\documentclass[12pt,preprint]{aastex}
\usepackage[english]{babel}
\usepackage{array}
\usepackage{color}
\usepackage{colortbl}

\shorttitle{Lensing by absorbers}
\shortauthors{Brice M\'enard}

\def\MgII{MgII}
\def\d{\mathrm{d}}
\def\be{\begin{equation}}
\def\ee{\end{equation}}
\def\bea{\begin{eqnarray}}
\def\eea{\end{eqnarray}}

\def\e{et al.}
\def\N{\mathrm{N}}
\def\C{\mathrm{C}}

\begin{document}

\title{Detecting and interpreting statistical lensing by absorbers}

\author{Brice M\'enard}
\affil{Institute for Advanced Study, Einstein Drive, Princeton NJ 08540, USA}

\begin{abstract}
  We propose a method for detecting gravitational magnification of distant
  sources, like quasars, due to absorber systems detected in their spectra. We
  first motivate the use of metal absorption lines rather than Lyman-$\alpha$
  lines, then we show how to relate the observed moments of the source
  magnitude distribution to the mass distribution of absorbers.  In order to
  illustrate the feasibility of the method, we use a simple model to estimate
  the amplitude of the effect expected for MgII absorption lines, and show
  that their lensing signal might already be detectable in large surveys like
  the SDSS.  Our model suggests that quasars behind strong MgII absorbers are
  in average brightened by $-0.05$ to $-0.2$ magnitude due to magnification.
  One must therefore revisit the claim that, in magnitude limited surveys,
  quasars with strong absorbers tend to be missed due to extinction effects.
  In addition to constraining the mass of absorber systems, applying our
  method will allow for the quantification of this bias.
\end{abstract}


\section{Introduction}

During the last decade, gravitational lensing has become an invaluable tool
for constraining the mass of many different structures (exoplanets, stars,
galaxies, clusters, large scale structure, etc.) and has significantly
improved our knowledge of many of these systems. However, there have been very
few results for absorption line systems, which is unfortunate because these
objects present a number of interesting aspects: low-ionisation metal lines
and strong HI lines sample the cold dense gas bound in galactic systems,
independent of the galaxy luminosities. Moreover absorption-line measurements
are largely insensitive to redshift and give us
access to early stages of galaxy formation.\\
Gravitational lensing by absorbers has been investigated theoretically by a
number of authors. So far all studies have focused on hydrogen line absorbers,
which offer easier theoretical modeling than metal lines since complications
due to gas metallicities can be avoided.  For most of these investigations,
lensing has been studied for its effects on quasar luminosity functions, i.e.
the magnification bias, and on the observed distribution of impact parameters
with respect to the center of the lenses (Pei 1995, Bartelmann \& Loeb 1996,
Smette \e\ 1997, Hamana et al. 2000, Le Brun \e\ 2000, Maller et al. 2002,
Perna et al.  2002).  However, only a few authors have used this technique to
study the absorber systems themselves.  Several studies have used multiple
quasar images to probe the extent of gas clouds (Smette et al. 1995, Kobayashi
et al. 2002, Ellison et al 2004), but in the context of absorbers,
gravitational lensing has rarely been used to constrain the mass properties of
the lenses, \emph{i.e. the purpose for which it is usually used}.  As far as
we know, only Maller et al. (2002) have proposed a method to investigate the
properties of an absorber population using lensing. They suggested analyzing
the distribution of Lyman-limit systems as a function of column density,
assuming that the corresponding unlensed distribution follows a power law, and
they showed how to use the measured deviations to obtain a lower limit for the
mass to gas column density ratio of these systems.  Unfortunately, their
method requires a
very large number of quasars and has not been applied so far.\\

This general lack of observational results for absorber systems might
originate in the preferential focus of previous studies on Lyman-$\alpha$
absorbers. Indeed, analytic models can be used more easily than for other
absorption lines and strong systems like DLAs are expected to trace the inner
part of dark matter halos and therefore favor gravitational magnification
effects.  However the quantity relevant to obtaining observational results is
not simply the amplitude of lensing effects, but the signal-to-noise ratio of
the detection that can be achieved. In such a case, strong Lyman-$\alpha$
absorbers might not be optimal candidates because large samples are difficult
to build.

Whereas for damped Lyman-$\alpha$ absorbers theoretical studies exist but
observational results are lacking, the situation is the opposite for metal
absorbers: no theoretical prediction has been made so far but a few authors
have attempted to detect magnification effects of metal absorber systems.
The first observational approach was to fit lensing models to the redshift
evolution of quasar absorbers to disentangle intrinsic evolution from
gravitational lensing (Thomas \& Webster 1990, Borgeest \& Mehlert 1993);
however, these studies could not find any signal from lensing. Another
approach first suggested by York \e\ (1991) was to divide a sample of quasar
spectra into bright and faint subsamples in order to determine the incidence
of metal absorbers (CIV, SiIV) in each subsample separately. However these
authors found no clear evidence of gravitational lensing.  Van den Berk \e\ 
(1996) extended the analysis to a larger sample of quasar spectra compiled
from the literature.  They found an excess of CIV absorbers in luminous
quasars but did not find a similar trend in the available \MgII\ sample.  More
recently, M\'enard \& P\'eroux (2003) used the 2dF-Quasar
survey to compare the number of quasars with and without absorbers as a
function of magnitude.  They found a relative excess of bright quasars with
absorbers as well as indications of reddening effects.  Finally, M\'enard,
Nestor and Turnshek have performed a more detailed analysis with data from the
Sloan Digital Sky Survey (SDSS) and find similar results (M\'enard et al., in
preparation).\\
The variety of results obtained in all these analyses turn out to be rather
difficult to interpret because the absorber species, redshift and equivalent
width ranges differ.
Moreover, as emphasized by M\'enard \& P\'eroux (2003), such analyses can
easily be biased if the selection of the objects is not done carefully enough.
For example, since absorption lines are more easily detected in high
signal-to-noise ratio spectra and since these tend to correspond to bright
quasars, a naive absorption line selection would introduce systematic
magnitude shifts.  In addition, the lack of theoretical estimates did not
allow for any guidance in interpreting the existing observational results.

In this paper we present the first calculation of statistical magnification
effects due to metal absorbers and investigate the possible constraints on the
absorber mass distribution. In section \ref{section_theory} we first motivate
the use of metal absorption lines, which possess a number of properties better
suited for statistical analyses than Lyman-$\alpha$ lines.  Then we present a
method for extracting the signal expected from gravitational magnification and
show how to use it to constrain the mass distribution of these systems. In
section \ref{section_mu_pdf}, in order to demonstrate the feasibility of this
technique, we use recent data on MgII systems, along with a simple lensing
model to estimate the amplitude of the expected effects. Finally, we estimate
the corresponding noise level in section \ref{section_detectability} and show
that the SDSS might provide the necessary data for a detection.

\section{Detecting the statistical magnification of quasars}
\label{section_theory}

\subsection{The choice of absorption lines}
\label{section_choice}

Narrow metal absorption lines and damped Lyman-$\alpha$ lines found in quasar
spectra are usually believed to indicate the presence of a galaxy along the
line-of-sight (Bahcall \& Spitzer 1969). In some rare cases, in which the
absorbing galaxy is massive enough and the impact parameter is less than a few
kpc, such a source-lens-observer configuration can give rise to multiple
quasar images (see for example Turnshek et al. 1997).  However, most often
when absorption lines are detected in a quasar spectrum, such configurations
are not encountered and lensing modifies the magnitude of the source without
producing additional images. Measuring these lensing effects can then only be
done in a statistical sense, because we do not know a priori the intrinsic
magnitude of a given background quasar. A large sample of quasars and absorber
systems is
therefore required in order to detect the magnification due to absorbers.\\
As mentioned in the Introduction, gravitational lensing by strong
Lyman-$\alpha$ absorbers has already been investigated theoretically by a
number of authors, motivated by the fact that such absorption features trace
the inner part of galactic halos and are therefore expected to favor
gravitational lensing effects. Unfortunately, Lyman-$\alpha$ lines can only be
seen at redshifts $z\gtrsim2$ from the ground so that only a small fraction of
observed quasars can allow for their detection.  It is thus difficult to build
a large sample of such absorption lines. Another drawback related to the high
redshift ranges comes from the fact that the lensing efficiency drops if both
lenses and sources are located at $z\gtrsim2$.

The statistical nature of the problem and the redshift ranges involved
motivate the use of metal lines. Given the large range of possible absorption
wavelengths they can be found at various redshifts from the ground, they can
therefore be selected in order to maximize the lensing efficiency, and they
probe a large range of galactic impact parameters, from a hundred kpc down to
a few kpc.  Therefore metal lines appear to be more appropriate than
Lyman-$\alpha$ lines for the purpose of detecting statistical magnification
effects. The method presented below can be applied to absorption lines like
MgII, FeII, SiII, CIV, etc. In section \ref{section_mu_pdf} we will explore
its feasibility by estimating the signal expected in the case of MgII
absorption lines.  Their 2796,2803 $\mathrm{\AA}$ doublet can be observed from
the ground over a very large range of redshifts ($0.2\le z \le 2.2$), they are
astrophysically abundant, and they have been seen to be very close tracers of
optically thick H I gas (Bergeron \& Stasinska 1986). We will consider strong
MgII absorbers, i.e. with a rest equivalent width
$W_0\gtrsim0.3\;\mathrm{\AA}$. Such systems can be detected with the
relatively low spectroscopic resolutions $(\mathrm{R}\sim 2000)$ used in
current large surveys like 2dF and SDSS.

\subsection{Extracting the signal}

The presence of a galaxy close to the line-of-sight of a background source can
modify its brightness in two ways: first, it can act as a gravitational lens
and amplify the flux of the source and second, the presence of dust around the
galaxy can extinct and redden the source's light. This can be described by
\begin{equation}
\mathrm{f}(\lambda)=\mathrm{f}_\mathrm{ref}(\lambda)\,
e^{-\tau(\frac{\lambda}{1+z_\mathrm{a}})}\times \mu ~,~~~~~~~~~~~~~~~
\end{equation}
where $\mathrm{f}(\lambda)$ is the altered flux of a source behind an galaxy,
$\mathrm{f}_\mathrm{ref}(\lambda)$ is the flux that would be observed without
intervening system, $\tau_\lambda$ is the optical depth of the galaxy and
$\mu$ its gravitational amplification.  The corresponding magnitude change can
then be written
\bea
\delta m &=& m-m_{\mathrm{ref}}\nonumber\\
&=&-2.5\,\log(\mu)+\frac{2.5}{\log e}\,\tau(\frac{\lambda}{1+z_\mathrm{a}})\;.
\eea 
Whereas gravitational magnification is achromatic, extinction effects strongly
depend on wavelength and usually become small in near infrared bands.  In this
paper we will focus on magnification only, arguing that observations at
appropriate wavelengths can isolate the lensing effects. Below, it will be
shown how to get constraints on dust extinction of metal absorbers using the
proposed measurements.

Let us consider an area of the sky which is large enough for the mean
magnification to be close to unity. In this area, let us consider a population
of sources, with a fraction of them being lensed.  Let $\mathrm{N_{ref}}(m)$
be the intrinsic magnitude distribution of the the unlensed sources and let
$\mathrm{N_{lens}}(m)$ be that of the lensed ones.  Let $f$ be the fraction of
sources behind a lens and $\mathrm{P}(\delta m)$ be the distribution of
magnitude shifts induced by these systems. We then have the following relation
between the magnitude number counts:
\begin{equation}
\N_\mathrm{lens}(m)=f\times\int \N_\mathrm{ref}(m-\delta m)
\;\mathrm{P}(\delta m) \; \d(\delta m)~.
\label{eq_convolution}
\end{equation}
If the lenses can be counted independently from the sources, $f$ can then be
estimated separately and information on the magnification properties of the
lenses, i.e. the distribution $\mathrm{P}(\delta m)$, can be extracted from
the observed number of lensed sources at a given magnitude.  This information
has been the one measured for example in the case of quasar-galaxy
correlations induced by lensing. In such a case, the magnification bias
changes the number density of background sources angularly close to foreground
lenses (see for example Bartelmann \& Schneider 2001).  However, if lenses
cannot be observed independently from their background sources, as is the case
for systems seen in absorption, a degeneracy exists between the intrinsic
number of lenses and the amplitude of the magnification bias, at a given
magnitude.  As shown in Eq.  \ref{eq_convolution}, increasing the fraction of
lenses $f$ or increasing their magnification, i.e. $-\delta m$, produce
similar effects at a given magnitude. Therefore, probing lensing effects must
be done by measuring changes in the shape of the magnitude distribution of the
lensed population, $\mathrm{N_{lens}}(m)$, with respect to that of the
reference population $\mathrm{N_{ref}}(m)$.  In the following we will estimate
this effect in the context of absorption lines detected in quasars spectra.

Let us consider a population of quasars with absorber systems having a rest
equivalent width greater than a certain threshold $W_{min}$. Let
$\mathrm{N_{lens}}(m)$ be the intrinsic magnitude distribution of this quasar
population, and let $\N_\mathrm{ref}(m)$ be that of a population of random
quasars. It should be noted that lensing effects by absorber systems with
equivalent widths smaller than $W_{min}$ also occur.  However, they are
expected to statistically affect the two distributions $\N_\mathrm{ref}(m)$
and $\mathrm{N_{lens}}(m)$ in the same way and will therefore not introduce
any change between them.

The convolution introduced in Eq. \ref{eq_convolution} implies that the
distribution of induced magnitude changes $\mathrm{P}(\delta m)$, which
contains information on absorber masses, can in principle be recovered via
Fourier space or using appropriate deconvolution techniques.  However a
number of constraints complicate such analyses in practice. The quantities
directly accessible to observations are
\bea
\N_\mathrm{ref}^\mathrm{obs}(m)&=&\N_\mathrm{ref}(m)\,
\mathrm C(m)\nonumber\\
\N_\mathrm{lens}^\mathrm{obs}(m)&=&\N_\mathrm{lens}(m)\,
\mathrm C(m)\;,
\label{eq_fourier}
\eea
where $\mathrm C(m)$ is the completeness of the detection procedure. For a
purely flux-limited sample of point sources, we have $\mathrm C(m)=1$ if
$m<m_{limit}$, where $m_{limit}$ is the survey limiting magnitude, and 0
otherwise. For quasars, however, the situation is different.  Their selection
depends on both their magnitude and colors. In practice, the corresponding
completeness functions usually drop from unity to zero over some magnitude
range that is significant compared with the magnitude range of the survey.
Recovering the function
$\N_\mathrm{lens}(m)=\N_\mathrm{lens}^\mathrm{obs}(m)/\mathrm C(m)$ will
therefore be very noisy in the range of magnitudes where $\mathrm C(m)$
becomes small.  Here we propose to extract the lensing signal by using the
moments of the observed magnitude number counts, and use them in order to
constrain the distribution of induced magnitude changes $\mathrm{P}(\delta
m)$.  Let us first define the observed mean magnitudes of the two quasar
populations:
\begin{equation}
\left\langle{m_\mathrm{ref}}\right\rangle=
\frac{ \int \N_\mathrm{ref}^\mathrm{obs}(m)\;m\;\d m}
{\int \N_\mathrm{ref}^\mathrm{obs}(m)\;\d m}~~;~~
\left\langle{m_\mathrm{lens}}\right\rangle=
\frac{ \int \N_\mathrm{lens}^\mathrm{obs}(m)\;m\;\d m}
{\int \N_\mathrm{lens}^\mathrm{obs}(m)\;\d m}~.
\label{eq_first_moment}
\end{equation}
These quantities are directly obtained from observations and do not
require binning the data. We can now define the \emph{observable} mean
magnitude shift as:
\begin{equation}
\Delta\langle{m_\mathrm{obs}}\rangle=
\left\langle{m_\mathrm{lens}}\right\rangle-
\left\langle{m_\mathrm{ref}}\right\rangle~.
\label{eq_obs}
\end{equation}
Similarly, we can introduce higher-order moments: 
\bea
\left\langle m_{\mathrm{ref}}^i\right\rangle&=& \frac{\int
  \N^\mathrm{obs}_\mathrm{ref}(m)\;(m-\left\langle{m_\mathrm{ref}}\right\rangle)^i\;\d
  m}
{\int \N^\mathrm{obs}_\mathrm{ref}(m)\;\d m}\nonumber\\
\left\langle{m_{\mathrm{lens}}^i}\right\rangle&=& \frac{\int
  \N^\mathrm{obs}_\mathrm{lens}(m)\;(m-\left\langle{m_\mathrm{lens}}\right\rangle)^i\;\d
  m} {\int \N^\mathrm{obs}_\mathrm{lens}(m)\;\d m}~,
\label{eq_higher_order_moments}
\eea
and the corresponding observable differences:
\begin{eqnarray}
\Delta\langle m_{\mathrm{obs}}^i \rangle=
\left\langle{m_{\mathrm{lens}}^i}\right\rangle-
\left\langle{m_{\mathrm{ref}}^i}\right\rangle
\label{eq_obs_higher}
\end{eqnarray}

As we can see by combining the previous equations, the observable quantities
$\Delta\langle{m_{\mathrm{obs}}^i}\rangle$ can be used to probe
the unknown distribution $\mathrm{P}(\delta m)$. We have for example for the
first-order moment:
\begin{eqnarray}
\Delta\langle{m_\mathrm{obs}}\rangle&=&
\left\langle{m_\mathrm{lens}}\right\rangle-
\left\langle{m_\mathrm{ref}}\right\rangle\nonumber\\
&=&\frac{ \int\d m\;\C(m)\;m\; \int \d(\delta m)\;
  \N_\mathrm{ref}(m-\delta m)\;\mathrm{P}(\delta m)}
{ \int\d m\;\C(m)\; \int \d(\delta m)\; \N_\mathrm{ref}(m-\delta
  m)\;\mathrm{P}(\delta m)}
-\frac{ \int \C(m)\;m\;\N_\mathrm{ref}(m)\;\d m}
{\int \C(m)\;\N_\mathrm{ref}(m)\;\d m}~.
\label{eq_all}
\end{eqnarray}
Since $\mathrm{N_\mathrm{ref}}(m)$ can be measured from the general
distribution of quasars without absorbers and $\mathrm{C}(m)$ can be estimated
for a given survey, this equation shows that a model for the distribution of
induced magnitude changes $\mathrm{P}(\delta m)$ can be constrained from
observations.  Eq. \ref{eq_all} can be generalized to higher-order moments.
It is then possible to define the likehood ${\cal L}(P)$ of the $n$ measured
moments $\Delta\langle m_\mathrm{obs}^{i=1,n}\rangle_\mathrm{data}$ and
minimize it in order to find the optimal parameters for $\mathrm{P}(\delta
m)$, i.e. a quantity related to the distribution of absorber masses and impact
parameters. A Kolmogorov-Smirnov test can also be used in order to compare the
observed and predicted distributions $\N_\mathrm{lens}(m)$, and find the
optimal parameters for $\mathrm{P}(\delta m)$.
    
The \emph{observed}-magnitude changes strongly depend on the slope of the
quasar luminosity distribution. Indeed the final effect can either be an
increase or a decrease of the mean magnitude of quasars with absorbers
relative to those without absorbers. Furthermore, magnification effects
\emph{cannot} be observed if the number of quasars as a function of luminosity
follows a power law. Indeed, in this case we would have
$\N_\mathrm{ref}(m)\propto a^m$ and, as can be seen from Eq.
\ref{eq_first_moment} and \ref{eq_obs}:
\be
\N_\mathrm{lens}(m)\propto \int a^{(m-\delta m)}
\;\mathrm{P}(\delta m) \; \d(\delta m) \propto a^m~.
\ee
Therefore the moments of $\N_\mathrm{lens}(m)$ and $\N_\mathrm{ref}(m)$ are
the same and $\Delta\langle m_{\mathrm{obs}}^i\rangle=0$, whatever the value
of the induced magnitude shift $\delta m$ and $\C(m)$.
This absence of observational lensing effects can be understood in the
following way: gravitational magnification makes each source brighter but
allows also new sources to be detectable as they become brighter than the
limiting magnitude of a given survey. This latter effect increases the number
of faint sources and tends to diminish the mean flux of the detected objects.
For a power law luminosity distribution, the flux increase due to
magnification effects is exactly canceled by the additional faint sources that
enter the sample due to magnification, and lensing effects due to absorbers
cannot be observed.  Therefore, it is possible to detect changes in the
magnitude distribution of a population lensed by absorbers only if its
unlensed number counts depart from a power law as a function of luminosity.

It is interesting to note that, on the observational side, constraints on dust
extinction can be obtained by measuring the moment excess in different
bands: the quantities $\Delta\langle{m_\mathrm{obs}^{i=1...n}}\rangle_{\lambda}
-\Delta\langle{m_\mathrm{obs}^{i=1...n}}\rangle_{\lambda'}$ where $\lambda$
and $\lambda'$ are two different wavelengths, will directly probe dust
extinction and will not be sensitive to magnification effects. This will allow
us to probe the statistical properties of the reddening curve of absorbers
systems as well as the relative optical depth as a function of absorber rest
equivalent width (M\'enard et al., in preparation). 

In the next section we will quantify the detectability of statistical lensing
effects of quasars by absorbers. We will first estimate the expected
distribution of induced magnitude changes due to a population of MgII
absorbers and then compute the expected excess in the moments of the observed
magnitude distributions $\Delta\langle{m_{\mathrm{obs}}^i}\rangle$.

\section{Magnification probability distribution function}
\label{section_mu_pdf}

\subsection{Induced magnitude changes}
\label{subsection_mu_pdf}

In the following we estimate the magnification effects of absorber systems
by assuming a mass profile and using existing observations of the distribution
of absorber impact parameters. Our analysis does not aim at presenting an
accurate model for absorber systems but only at giving a rough estimate of their
expected magnification effects and at showing how to use them to constrain the
mass distribution of absorbing galaxies.

\subsubsection{Absorber model}

As mentioned in section \ref{section_choice}, we focus only on MgII absorption
lines that can be detected in current large surveys, i.e. with $W_0\gtrsim0.3~
\mathrm{\AA}$.  Based on the 2dF Quasar Survey, Outram et al (2001) have
detected about one hundred such absorbers in the spectra of roughly one
thousand quasars with high signal-to-noise ratio and only about one percent of
quasars with two strong absorbers in their spectrum.  This shows that multiple
lensing of a single quasar can safely be neglected in our case.\\
In the following, we will assume for convenience that the mass distribution of
the galaxies responsible for the absorption is described by a singular
isothermal sphere profile (more sophisticated magnification calculations are
presented in Perrotta \e\ 2002).  The galactic halos therefore have the
following density and surface density:
\be
\rho \propto \frac{\sigma_v^2}{r^2}~~~~\mathrm{and}~~~~~
\Sigma(r)=\frac{\sigma_v^2}{2\mathrm{G}r}\;,
\label{eq_sis}
\ee
and the corresponding magnification effects of a point source can then be
computed in a simple way:
\begin{equation}
  \mu = \left\{
  \begin{array}{ll}
    {2/y} & \hbox{if $y\le1$} \\
    1 + {1/y} & \hbox{if $y\ge1$} \\
  \end{array}\right.
\label{eq_mu}
\end{equation}
where $y$ is the impact parameter normalized to the Einstein radius of the
lens: 
\begin{equation}
  \zeta_0 \equiv 4\pi\left(\frac{\sigma_v}{c}\right)^2\,
  \frac{D_{\rm d}\,D_{\rm ds}}{D_{\rm s}}\;,
\label{eq:10}
\end{equation}
where $\sigma_v$ is the velocity dispersion and $D_{\rm d,s,ds}$ are the
angular diameter distances from the observer to
the lens, to the source, and from the lens to the source.

\begin{figure}[h]
\begin{center}
  \includegraphics[height=6cm,width=7cm]{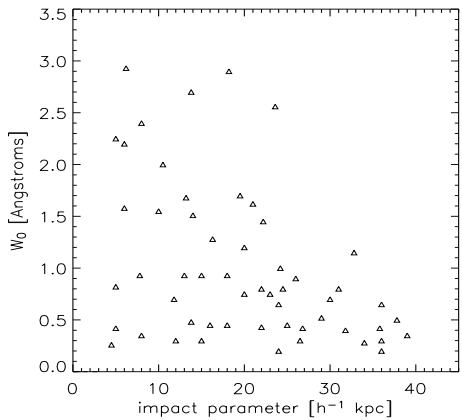}
  \caption{Distribution of rest equivalent widths $W_0$ as a function of
    impact parameter between quasars and the center of absorbing
    galaxies, measured by Steidel \e (1994).}
\label{plot_Steidel}
\end{center}
\end{figure}

As can been seen from the above equations, the use of magnification effects to
constrain masses requires the knowledge of the relevant halo radii.  In the
case of MgII systems such quantities have been measured by Steidel \e\ (1994)
and Steidel (1995). They used deep observations of a sample of 58 MgII
absorber systems (with a mean redshift $z=0.6$) in addition to a spectroscopic
follow up in order to identify the galaxies responsible for the absorption.
From these observations they measured the distribution of absorber rest
equivalent widths $W_0$ as a function of galactic impact parameter $b$ (see
Fig.  \ref{plot_Steidel}). Below we will present different ways of using this
information in order to compute the magnification effects.

\subsubsection{Characteristic mass and magnification}

\begin{figure}[t]
\begin{center}
  \includegraphics[height=6cm,width=.44\hsize]{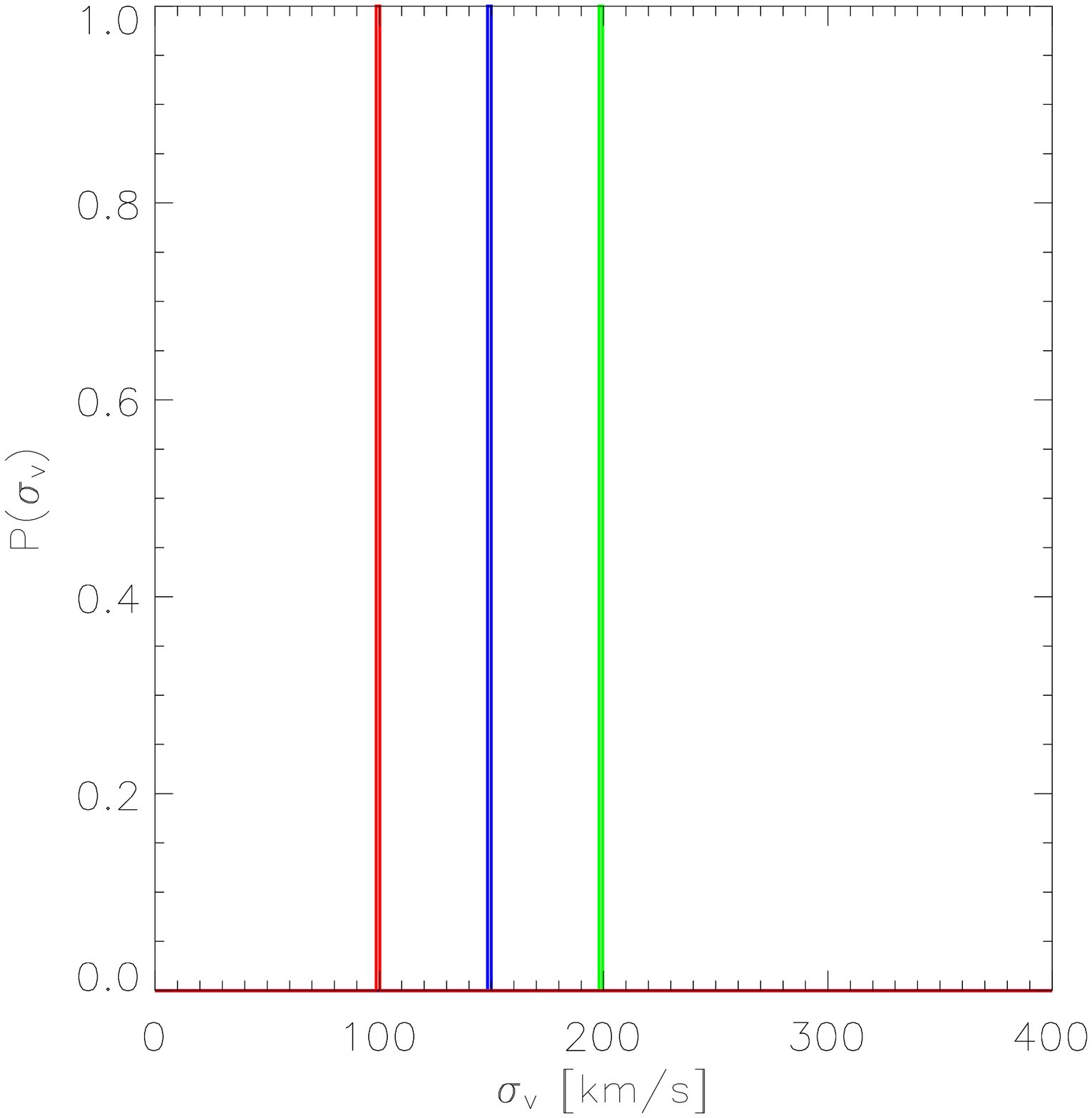}
  \includegraphics[height=6cm,width=.44\hsize]{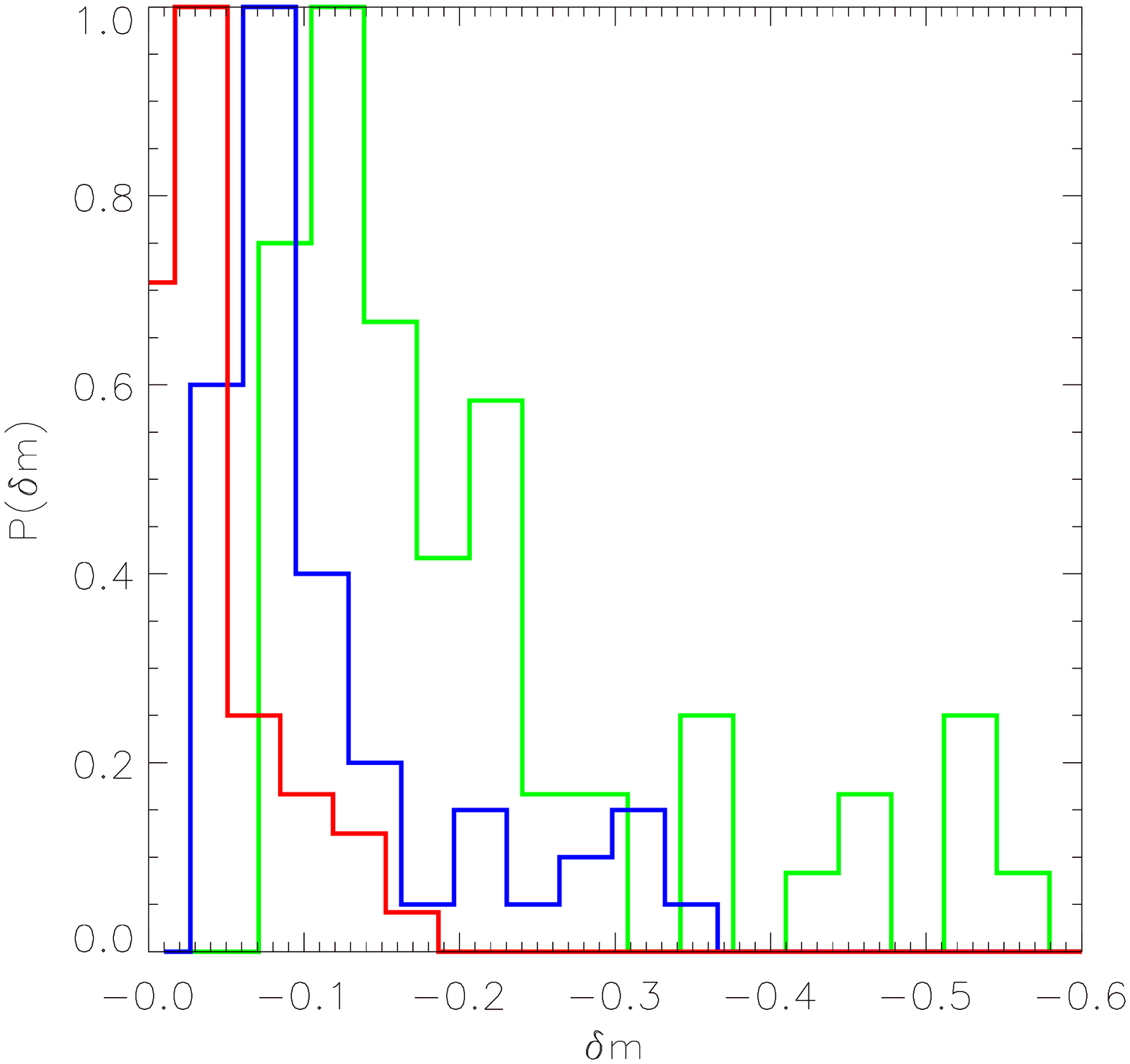}
  \caption{\emph{Left panel:} velocity dispersions characterizing the MgII
    absorbing galaxies. \emph{Right panel:} distributions of
    quasar magnitude shifts induced by these absorbers, considering the
    distribution of impact parameters measured by Steidel \e\ }
\label{plot_pdf_delta_m0}
\end{center}
\end{figure}

Using the distribution of impact parameters, we can first compute the
characteristic magnification expected for a given absorber mass.  To do so, we
will consider that all absorbing galaxies are identical and we will
investigate the effects of isothermal spheres with $\sigma_v=100, 150$ or
$200$ km s$^{-1}$. Considering all absorbers to be at redshift $z=0.6$, in
order to match the redshifts used by Steidel \e, we can then compute the
distribution of induced magnitude shifts $\delta m=-2.5\,\log(\mu)$. The
results are shown in Fig.  \ref{plot_pdf_delta_m0}. The left panel indicates
the different velocity dispersions considered and the right panel shows the
corresponding distributions of induced magnitude shifts. As can be seen, if
MgII absorbers are typically surrounded by galactic dark matter halos
represented by singular isothermal spheres, our calculations show that such a
population will change the mean magnitude of their background quasars by
$-0.02$ to $-0.2$ typically.  Such changes translate into quasar flux
increases of about 2 to 20\%. However, we recall that these induced magnitude
changes are not directly observable.  The observed magnitude changes depend on
the shape of the luminosity distribution of the sources and will be computed
in the section \ref{section_obs_changes}.

\subsubsection{Gas follows dark matter}

\begin{figure}[t]
\begin{center}
  \includegraphics[height=6cm,width=.44\hsize]{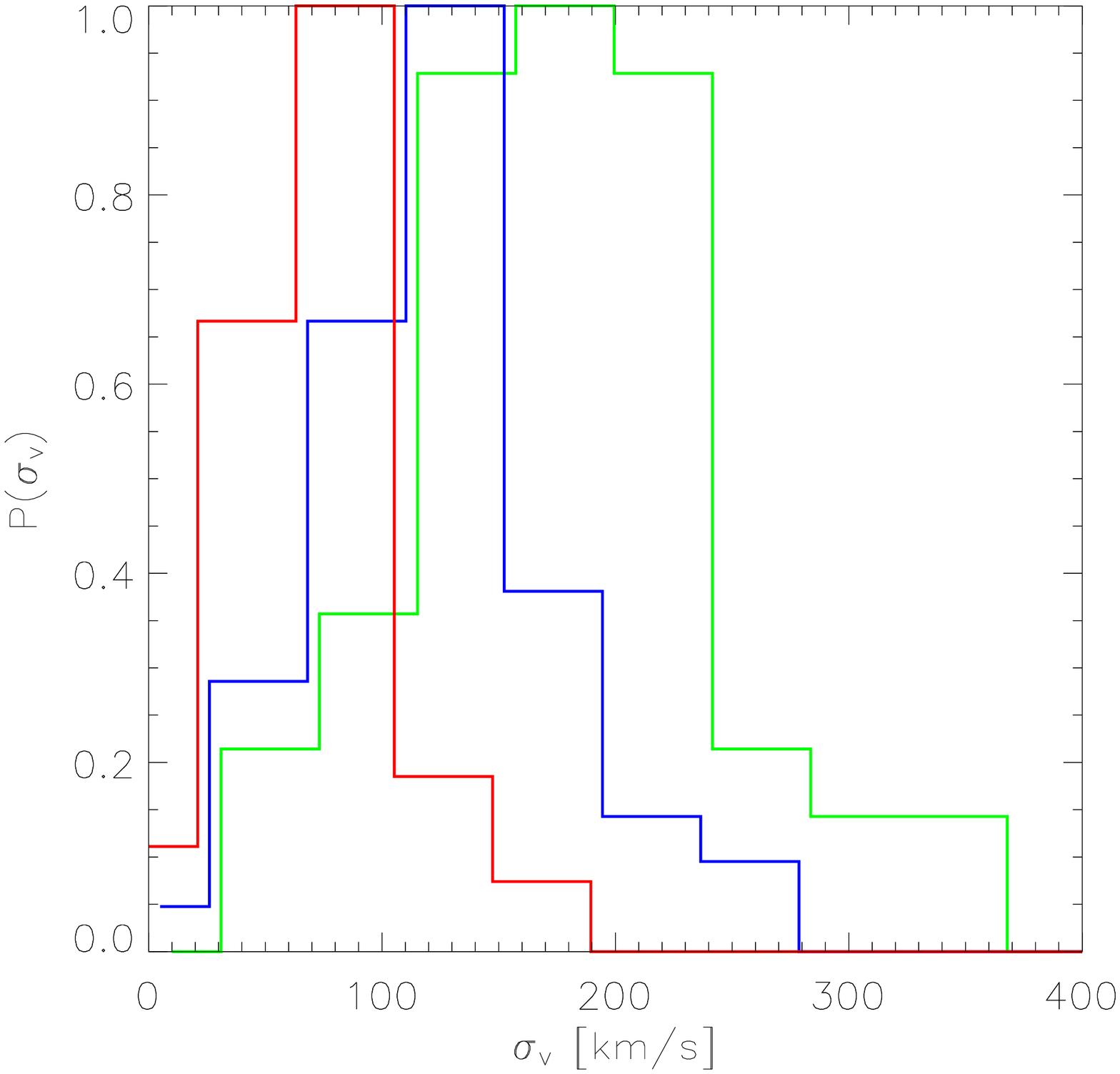}
  \includegraphics[height=6cm,width=.44\hsize]{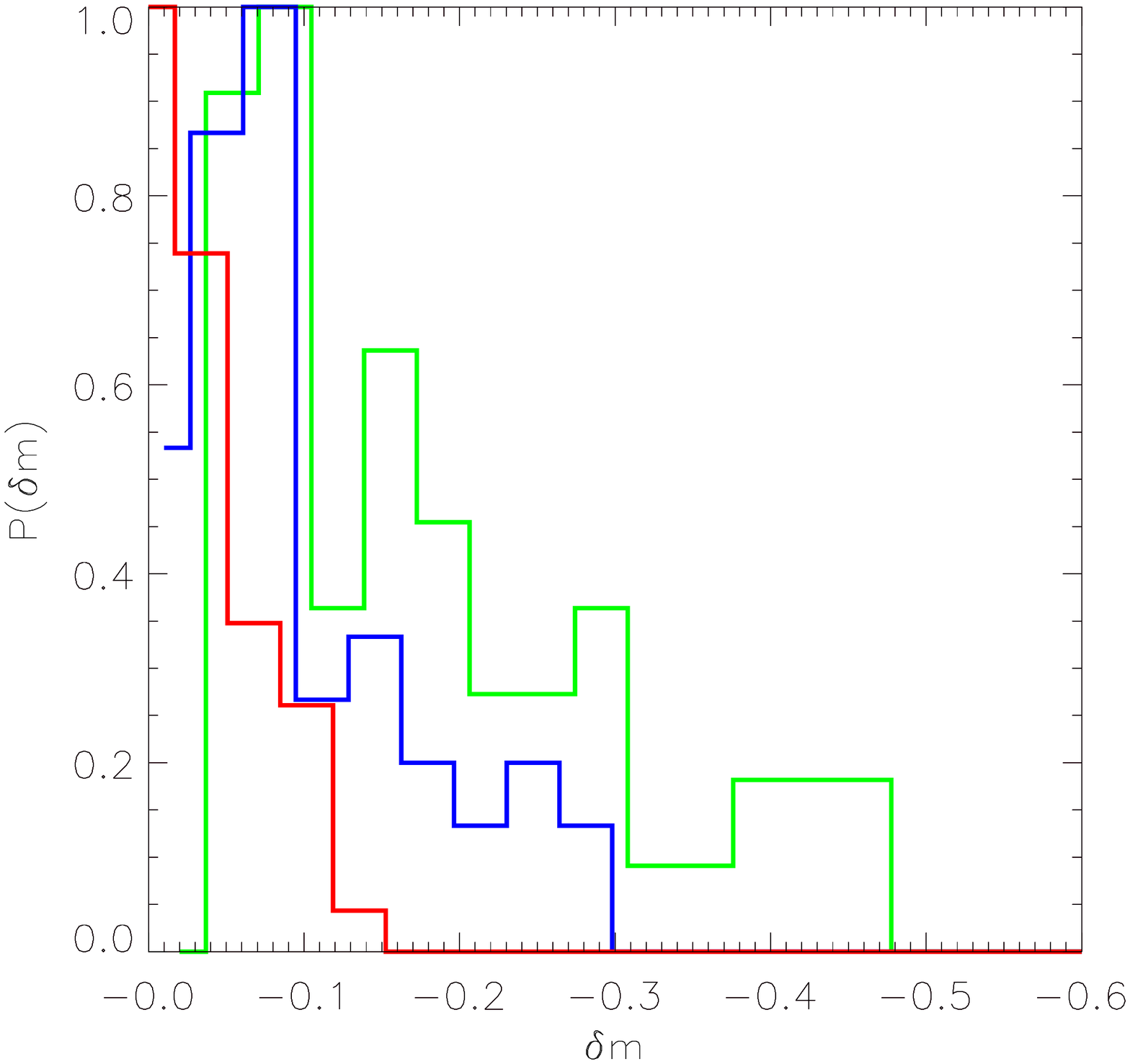}
  \caption{\emph{Left panel:} distribution of velocity dispersions characterizing the MgII
    absorbing galaxies obtained by assuming a deterministic relation between
    gas and mass. \emph{Right panel:} distributions of quasar magnitude shifts
    induced by these absorbers, as for Fig. \ref{plot_pdf_delta_m0}}
\label{plot_pdf_delta_m1}
\end{center}
\end{figure}

The distribution of absorber rest equivalent widths as a function of impact
parameter observed by Steidel \e\ (Fig. \ref{plot_Steidel}) reveals a
correlation at the $3.8\sigma$ level between these two parameters. Such an
information can be taken into account in our modelling of the magnification
effects. Let us first assume that the gas distribution is proportional to that
of the dark matter. In this case we have:
\be
\mathrm{N_{MgII}}(b)= \mathrm{f_{MgII}}(b)\times\Sigma(b)
\label{eq_assumption1}
\ee
where $\Sigma$ is the total mass surface density, and $\mathrm{f_{MgII}}$ is
the MgII to dark matter fraction, i.e. a quantity that depends on the baryon
fraction, the metallicity and the ionization background.\\

It is useful to note that, as infered by Bahcall (1975), the MgII gas
distribution within a galactic halo is actually due to a number of individual
clouds. This was recently verified by Churchill \e\ (1996). Using very high
spectroscopic resolution, these authors could detect the absorption lines of
individual clouds and they have shown that the overall rest equivalent width
$W_0$ is proportional to the number of individual velocity components
responsible for the global absorption feature.  Therefore, assuming that these
individual clouds have similar properties we have roughly $W_0 \propto
N_{MgII}$.  In this case, the $\mathrm{W_0}(b)$ distribution given by Steidel
\e\ (1994) can then be used to infer a distribution proportional to that of
the mass surface densitites:
\begin{equation}
\delta_\mathrm{D}\left[\mathrm{W_0}(b)\right] 
\Leftrightarrow \delta_\mathrm{D} \left[\Sigma(b)/\Sigma_0\right]
\Rightarrow \delta_\mathrm{D} \left[\sigma_v/\sigma_{v,0}\right]\,,
\label{eq_delta}
\end{equation}
i.e. a given value of rest equivalent width at an impact parameter $b$ maps
into a given value of velocity dispersion for which the normalisation
$\sigma_{v,0}$ is not known. In this case, the distribution plotted in Fig.
\ref{plot_Steidel} provides us with an estimate of the distribution of
relative velocity dispersions.  For a given normalisation, this distribution
can then be used to estimate the expected magnification effects.  In order to
do so, we convert each data point $\mathrm{W_0}(b)$ given by Steidel \e\ into
as estimate of relative velocity dispersions $\sigma_v/\sigma_{v,0}(b)$. We then
normalize the distribution of velocity dispersions such that
$\langle\sigma_v\rangle=100, 150$ or $200$ km s$^{-1}$.  The left panel of
Fig. \ref{plot_pdf_delta_m1} shows the infered distributions of $\sigma_v$ and
the right panel shows the corresponding distributions of induced magnitude
shifts $\mathrm{P}(\delta m)$ for absorbers at redshift $z=0.6$.  The changes
in the magnitudes of the background quasars are similar to the previous case.
They range typically from $-0.02$ to $-0.3$.

\subsubsection{Scatter in the gas-dark matter relation}
\label{subsection_stochasticity}

\begin{figure}[t]
\begin{center}
  \includegraphics[height=6cm,width=.44\hsize]{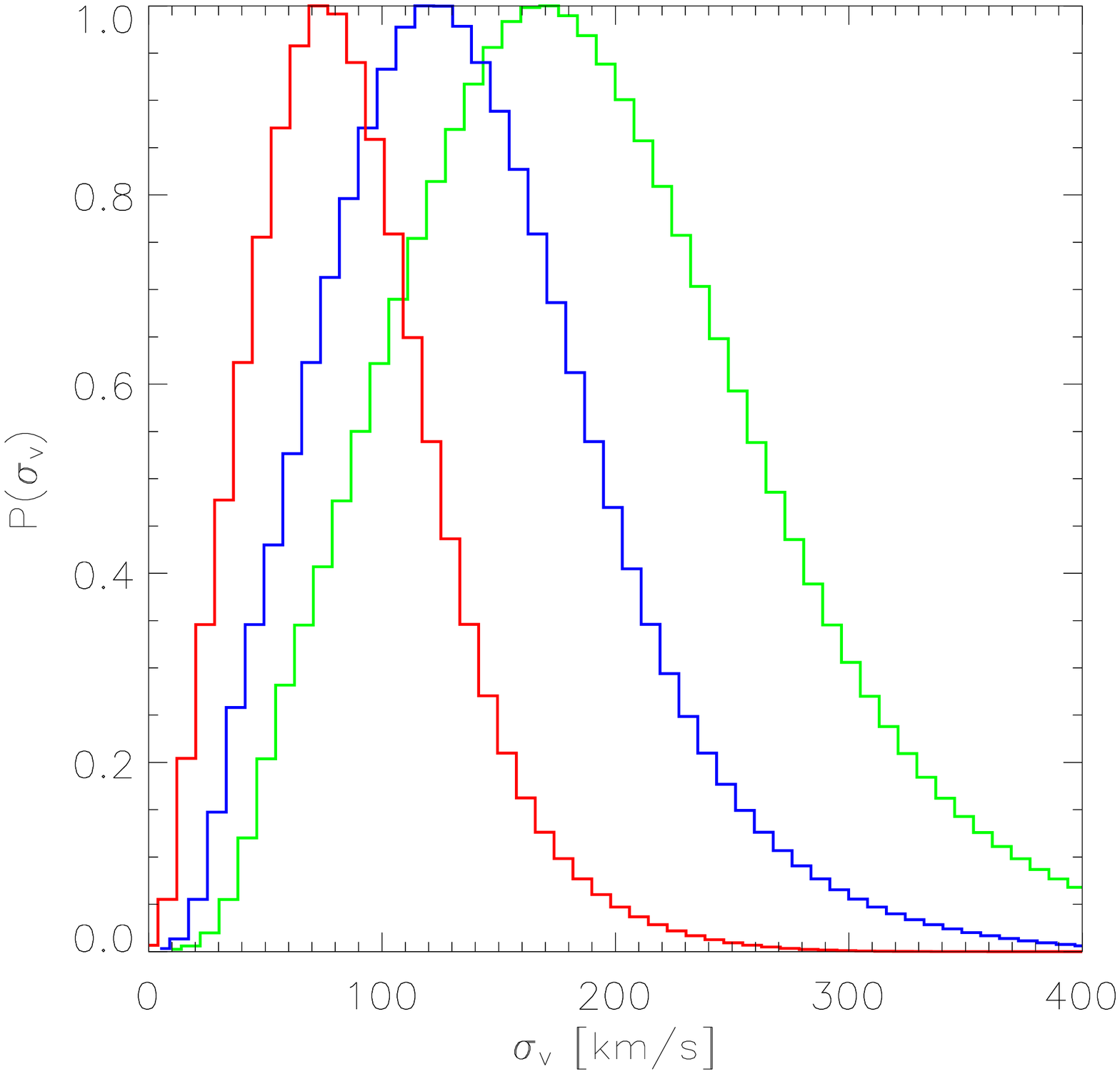}
  \includegraphics[height=6cm,width=.44\hsize]{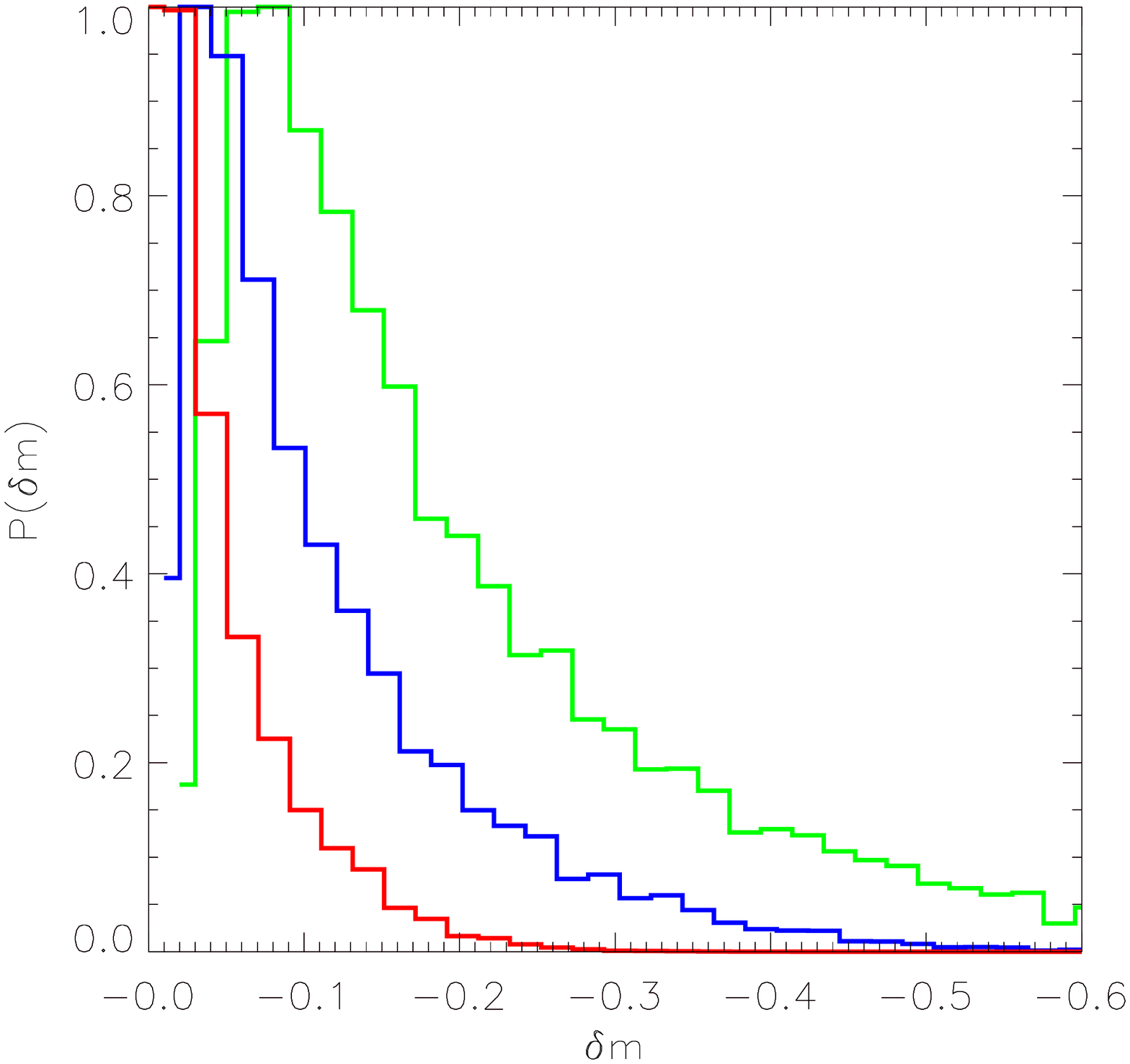}
  \caption{idem as Fig \ref{plot_pdf_delta_m0} but adding stochasticity in the
    gas-mass relation using Eq. \ref{eq_gaussian}. The width of the
    distribution $\mathrm{P}(\delta m)$ has increased compared that of Fig
    \ref{plot_pdf_delta_m0}}
\label{plot_pdf_delta_m2}
\end{center}
\end{figure}
In reality, a number of stochastic effects can alter the proportionality
between gas and mass distributions, and thus introduce some scatter in the
previous relation.  For instance, the assumption that all the individual
clouds giving rise to the overall absorption feature are similar might not
hold, or as it has been argued by Bond \e\ (2001), superwinds arising in
actively star-forming galaxies can give rise to strong MgII absorption lines
seen in some quasar spectra. Therefore it might be more realistic to allow for
some scatter in the relation between mass and gas. In order to do so, we can
introduce a distribution of velocity dispersions that can give rise to a given
gas column density. For example we can use:
\begin{equation}
\delta_\mathrm{D}\left[\mathrm{N_{MgII}}\right] 
\Rightarrow \frac{1}{s\,\sqrt{2\pi}}\;
\mathrm{exp}{\left(\frac{-\sigma_v^2}{2\,s^2}\right)} 
\label{eq_gaussian}
\end{equation}
where $s$ is the scatter in absorber velocity dispersions. Other relations can
be used to increase or decrease the scatter of mass with respect to that of
the gas.  As mentioned previously, observational constraints from higher-order
moments will be sensitive to such properties and can allow us to test the
validity of a given model.  In order to illustrate this idea, we have computed
the distribution of induced magnitude changes for the same mean velocity
dispersions used in the previous section and we have added a scatter in the
absorber velocity dispersion of $\sigma_v/4$, where $\sigma_v$ is the mean
velocity dispersion.  A given value $W_0$ now maps into a distribution of
systems spanning a factor $\sim3$ in masses. The corresponding magnitude
changes are shown in the right panel of Fig.  \ref{plot_pdf_delta_m2}.  The
width of the distribution of induced magnitude changes $\mathrm{P}(\delta
m)$ has increased compared to the previous case where an exact proportionality
between gas and mass was assumed, and higher magnification values are reached.

\subsection{Observable magnitude changes}
\label{section_obs_changes}

As shown in section \ref{section_theory}, the observable magnitude changes
depend on the distribution of induced magnitude shifts $\mathrm{P}(\delta m)$,
described in the previous section, as well as the induced quasar magnitude
distribution $\mathrm{N_{ref}}(m)$ and the completeness function
$\mathrm{C}(m)$ of the relevant survey.\\
So far, the best estimate of the intrinsic quasar luminosity distribution has
been been obtained by Boyle \e\ (2000) and Croom \e\ (2004) using the 2dF
survey. These authors quantified the photometric and spectroscopic
imcompleteness of the survey and found that the quasar magnitude distribution
is well fitted by a broken power law, as measured by earlier authors. The
corresponding fitting formula for the differential number counts reads (Myers
et al. 2003):
\begin{equation}
\mathrm{N_{ref}}(m) = \frac{N_0}{10^{-\alpha(m-m_*)} + 10^{-\beta(m-m_*)}} 
\label{eq_completeness}
\end{equation}
with $m_*=19.1$, $\alpha=0.98$ and $\beta=0.15$.  In order to take into
account the completeness function, we will consider several forms. First, a
completeness corresponding to a purely magnitude limited survey:
\begin{equation}
  \mathrm{C_0}(m)= \left\{
    \begin{array}{ll}
      1 & \hbox{if $m<m_{limit}$} \\
      0 & \hbox{otherwise}, \\
    \end{array}\right.
\end{equation}
where $m_{limit}$ is the limiting magnitude of the survey. For quasars, the
color-based selection function introduces a more complicated completeness as a
function of magnitude and gives rise to a decrease of $\mathrm{C}(m)$ over
some magnitude range. In order to approximate the corresponding behavior we
will consider the two following cases:
\begin{equation}
  \mathrm{C_1}(m)= \left\{
    \begin{array}{ll}
      1 & \hbox{if $m<m_{b}$} \\
      \mathrm{e}^{-(m-m_b)} & \hbox{if $m_b<m<m_{limit}$} \\
      0 & \hbox{otherwise} \\
    \end{array}\right.
\end{equation}
\begin{equation}
  \mathrm{C_2}(m)= \left\{
    \begin{array}{ll}
      1 & \hbox{if $m<m_{b}$} \\
      \mathrm{e}^{-(m-m_b)^2} & \hbox{if $m_b<m<m_{limit}$} \\
      0 & \hbox{otherwise} \\
    \end{array}\right.
\end{equation}
In order to reproduce realistic observed magnitude distributions for quasars,
we will use $m_{limit}=20$ and $m_b=17.5$ in our numerical estimates.  Note
that in this case we have $\mathrm{C}(m)\approx0$ for $m\approx m_{limit}$.
Using $\mathrm{C}_0(m)$, $\mathrm{C}_1(m)$ and $\mathrm{C}_2(m)$ will allow us
to estimate how the magnitude changes are sensitive to the shape of the survey
completeness.

\begin{figure}[t]
\begin{center}
  \includegraphics[height=9cm,width=\hsize]{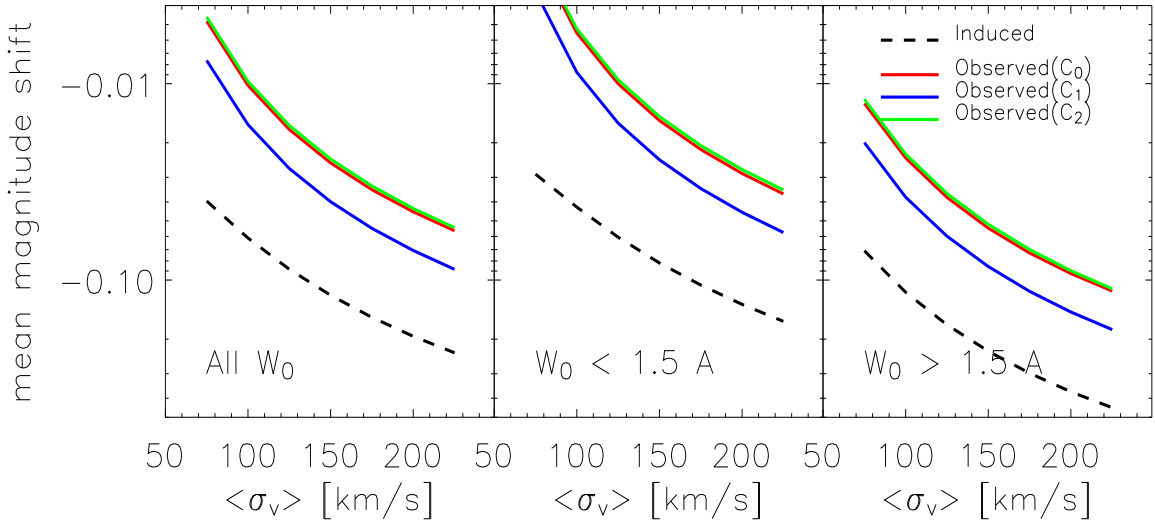}
  \vspace{-1.5cm}
  \caption{Mean magnitude shifts of quasars due to magnification by
    intervening MgII absorbers at $z=0.6$, as a function of the mean velocity
    dispersion of the absorbing galaxies. The dashed curve shows the real mean
    magnitude shift induced and the colored curves show the mean shift that is
    observed, as a result of the magnification bias. We have computed the
    signal expected for the three different completeness functions
    $\mathrm{C}(m)$ introduced in the text. The \emph{left} panel shows the
    effect for absorbers with $0.3<W_0<3.3~\mathrm{\AA}$. The \emph{middle}
    and \emph{right} panels show how the signal is expected to vary as a
    function of rest equivalent width $W_0$.}
\label{plot_first_order}
\end{center}
\end{figure}
All the elements are now in place to evaluate the observable shifts in the
moments of the magnitude distributions, i.e. evaluating Eq.
\ref{eq_first_moment} to \ref{eq_obs_higher}.  Fig. \ref{plot_first_order}
shows our computation of the mean magnitude shifts for the fiducial
populations of quasars and MgII absorbers, as a function of the mean absorber
velocity dispersion. We have plotted here the results obtained using the
deterministic relation between gas and mass. However, the values obtained for
the other scenarios, i.e. unique absorber mass and stochastic gas-mass
relation, are very similar and differ by less than $20\%$.  This similarity is
due to the fact that the magnification $\mu$ varies roughly linearly with the
mass in the weak lensing regime. We have: $\delta m\propto \log(\mu) \propto
\mu-1 \propto \sigma_v^2$. Therefore, for the first-order moments, the precise
relation between gas and mass does not strongly affect the results, and
$\Delta\langle m_\mathrm{obs}\rangle$ is primarily a measure of the average
mass of absorbers.  The left panel of the figure shows the magnitude shifts
expected when ones considers the whole range of rest equivalent widths
observed by Steidel \e , i.e.  $0.3<W_0<3.3~\mathrm{\AA}$. The dashed line
shows the mean induced magnitude shift $\langle\delta m\rangle$. It ranges
from $-0.04$ to $-0.20$ depending of the mean absorber velocity dispersion.
The colored lines show the corresponding observable mangnitude shifts
$\Delta\langle m_\mathrm{obs}\rangle$ for the three completeness functions
introduced above.  As can be seen, the observable effect is significantly
weaker than the real mean magnitude shift induced by the population of
absorbers. This is due to the fact that the quasar luminosity distribution
departs only weakly from a power law. The observable shift ranges from
$-0.005$ to $-0.07$ depending on the completeness function.  This signal can
be observed as a function of absorber rest equivalent width.  In the middle
and right panels of Fig \ref{plot_first_order} we have plotted the same
quantities for systems with $W_0<1.5~\mathrm{\AA}$ and $W_0>1.5~\mathrm{\AA}$.
We find that the mean observed magnitude shift $\Delta\langle m_{obs}\rangle$
is roughly three times larger for systems with $W_0>1.5~\mathrm{\AA}$ than
that of systems with $W_0<1.5~\mathrm{\AA}$.  Fig \ref{plot_first_order}
provides us with a way of recovering the real magnitude shift $\langle \delta
m \rangle$ from the observed one, $\Delta\langle m_\mathrm{obs}\rangle$.

\begin{figure}[t]
\begin{center}
  \includegraphics[height=6cm,width=.44\hsize]{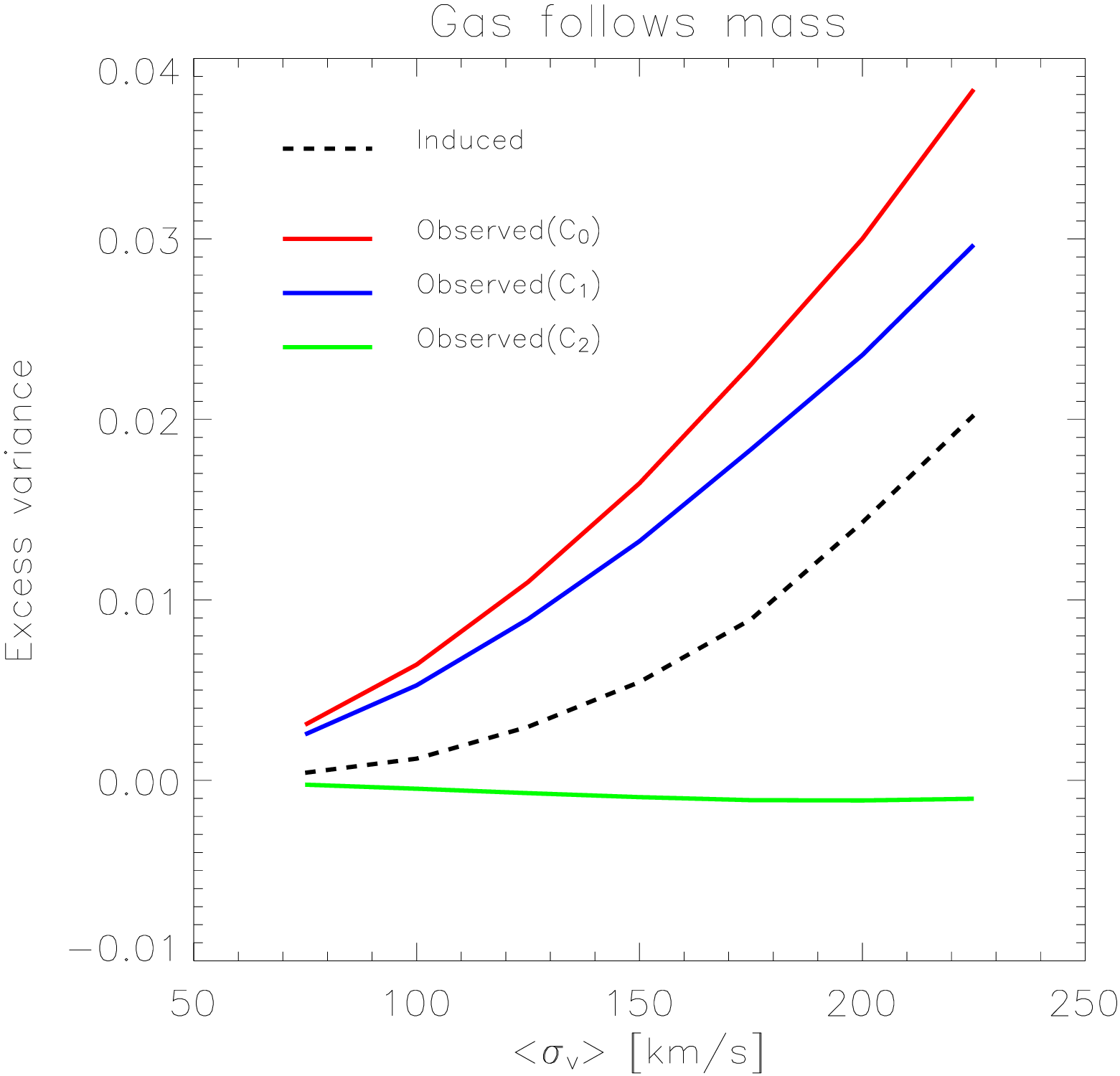}
  \includegraphics[height=6cm,width=.44\hsize]{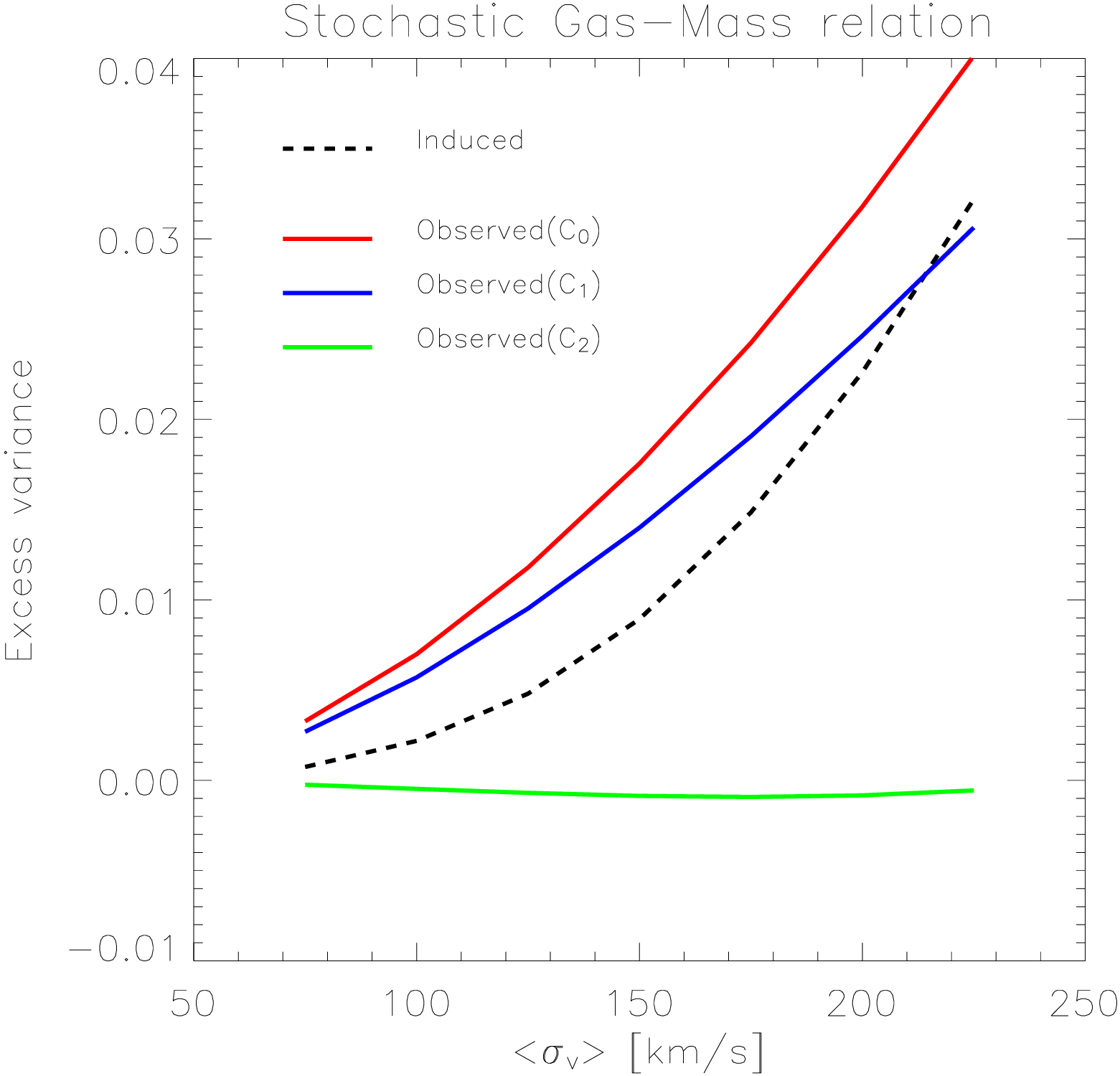}
  \caption{Excess variance of quasars magnitude due to magnification by
    intervening MgII absorbers. The dashed curve shows the real excess
    variance induced and the colored curves show the change in the magnitude
    variance that can be observed. In the left panel we have used a
    deterministic relation between gas and mass, and in the right panel some
    stochasticity has been added (see text).}
\label{plot_second_order}
\end{center}
\end{figure}

In Fig. \ref{plot_second_order} we show the results for the second-order
moments. The left panel displays the results for the deterministic gas-mass
relation. The real excess variance $\langle\delta m^2\rangle$ (dashed
line) varies from 0.001 to 0.02 depending on the mean absorber velocity
dispersion. The colored curves present the corresponding observable magnitude
changes $\Delta\langle m_{obs}^2\rangle$.  Whereas $\langle\delta m^2\rangle$
is always a positive quantity, it is interesting to notice that the observable
second-order moments changes $\Delta\langle m_{obs}^2\rangle$ can be negative
in some cases.  Indeed, given the existence of a limiting magnitude, shifting
the quasar magnitude distribution (Eq.  \ref{eq_completeness}) also results in
a change of its observed second-order moment. In some cases, depending on the
position of the magnitude break $m_b$ and the shape of the completeness
function, this results in a decrease of the observed magnitude variance. On
the other hand, the excess of variance introduced by the existence
of a distribution of magnifications tends to increase the observed variance of
magnitudes. The net effect can then be either positive or negative. For a
given value of $\langle\delta m^2\rangle$, $\Delta\langle m_{obs}^2\rangle$ is
expected to show a strong dependence on the shape of the completeness function
$\mathrm{C_i}(m)$.  The left panel of the figure shows that we expect
$\Delta\langle m_{obs}^2\rangle$ to lie between $-0.001$ and $0.04$.  The
right panel of the figure shows the same quantities for the stochastic
gas-mass relation, i.e. when a scatter of $\sigma_{v,0}/4$ in the velocity
dispersions is added. As expected, the induced magnitude variance is larger
than in the previous case.  However the changes in the observed second-order
moments are rather small compared to the deterministic case: the values of
$\Delta\langle m_{obs}^2\rangle$ differ only at the $\sim10^{-3}$ level.

\section{Detectability}
\label{section_detectability}

\begin{figure}[t]
\begin{center}
  \includegraphics[width=.4\hsize]{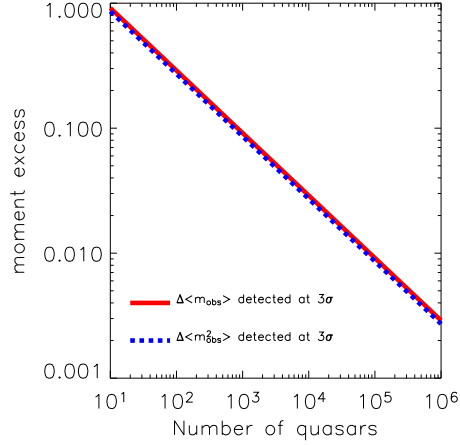}
  \caption{Number of quasars with absorbers required for detecting, at the
    3-$\sigma$ level, a change in the magnitude first- (solid line) and second-order
    (dashed line) moments. For this figure we have considered SDSS quasars observed
    in the $z$-band.}
\label{plot_nb_qso}
\end{center}
\end{figure}
Whereas our model has provided us with an estimate of the unavoidable
magnitude changes $\langle \delta m^i\rangle$ due to magnification, it does
not allow us to determine, in the general case, the detectability of
$\Delta\langle m_{obs}^i\rangle$. Including the effects of extinction by dust
would be required since they compete against magnification.  Unfortunately,
the statistical properties of dust extinction of MgII systems are still poorly
constrained. The following estimates for the detectability of $\Delta\langle
m_{obs}^i\rangle$ are therefore valid only if extinction effects can be
neglected, as is the case at large wavelengths.

The noise level related to the measurements of the magnitude moments depends
on the number of available quasars as well as the shape of the observed
magnitude distributions.  For a given distribution with centered moments
$\langle m^n\rangle$ of order $n$ and $N$ objects, the error on the estimation
of its mean reads
\begin{equation}
\sigma^2(\langle m\rangle)=\frac{\langle m^2\rangle}{N}
\end{equation}
and the error on the estimation of its variance reads
\bea
\sigma^2(\langle m^2\rangle)&=&\frac{N-1}{N^3}
\left[(N-1)\langle m^4\rangle-(N-3)\langle m^2\rangle^2\right]\nonumber\\
&\approx&\frac{1}{N}(\langle m^4\rangle-\langle m^2\rangle^2)
\eea

In order to estimate these quantities we will use the moments $\langle
m^2\rangle$ and $\langle m^4\rangle$ given by SDSS DR1 quasars (Schneider et
al 2003). Using the observed quasar magnitude distribution in $z$-band (where
extinction is weaker) we find: $\langle m^2\rangle\approx0.53$ and $\langle
m^4\rangle\approx0.75$.  Using these values, we show in Fig. \ref{plot_nb_qso}
the number of quasars required to observe a change in the magnitude first- and
second-order moments at the 3$\sigma$ level.  Considering that the mean
velocity dispersion of MgII absorber systems to be 150 km s$^{-1}$, our
previous calculations have shown that, $\langle\delta m\rangle\approx-0.15$
and $\langle\delta m^2\rangle\approx0.005$. Using the completeness function
$\mathrm{C_1}$, the observable changes are $\Delta\langle
m_{obs}\rangle\approx-0.1$ and $\Delta\langle m_{obs}^2\rangle\approx0.05$.
Figure \ref{plot_nb_qso} then shows that $\sim 2000$ systems are required to
detect a $3\sigma$ change in the mean magnitude of quasars, and $\sim 10\,000$
for an excess of the magnitude scatter.
A detection (or a non-detection) of such mean quasar magnitude shift will
provide us with some constraints on the mean mass of the absorbers. This
signal can be measured as a function of redshift and rest equivalent width.
Our estimate shows that the SDSS survey might already be able to provide us
with a detection of the magnitude shift due to gravitational lensing by MgII
absorbers.  The detection of an excess of variance in the magnitude
distributions might be possible.  However, the number of quasars required in
order to distinguish between a deterministic or stochastic relation between
gas and dark matter is several orders of magnitude larger. The corresponding
detection level will require next generation surveys.

\section{Discussion}

Investigating the statistical magnification effects of quasars by intervening
absorber systems presents several interests: our calculations have shown that
the mean observed magnitude shift $\Delta\langle m_{obs}\rangle$ provides us
with an estimate of the mean mass of absorbing galaxies, which turns out to be
weakly sensitive to the details of the relation between mass and gas.  In
addition to probing the mean mass of absorbing galaxies, measuring the mean
magnitude shifts $\Delta\langle m_{obs}\rangle$ will also allow us to quantify
the \emph{real} magnitude changes $\langle\delta m\rangle$ that are induced. This
will be of direct relevance for estimating the number of quasars missed due
extinction by dust or artificially added due to magnification, in magnitude
limited surveys.  We have shown that quasars behind absorbers are not
necessarily dimmed, as is usually mentioned in the literature, but they can
actually be brighter if gravitational magnification is more important than
extinction effects. By applying our method to MgII lines, we have shown that,
if MgII systems trace galactic halos with a mean velocity dispersion
$100<\langle\sigma_v\rangle<200$ km s$^{-1}$, the magnitude of their
background quasars will be changed by $-0.03$ to $-0.15$ for systems with
$0.3<W_0<1.5~\mathrm{\AA}$ and by $-0.1$ to $-0.4$ for systems with
$W_0>1.5~\mathrm{\AA}$ (see dashed line in Fig \ref{plot_first_order}).  In
this paper we have considered the absorbing galaxies to be isolated.  However,
these systems might be correlated to matter overdensities that do not
necessarily give rise to absorption lines. For example they might be located
in groups of galaxies. In these cases, the lensing effects might be
stronger than the ones derived in this paper.\\
Extinction effects due to high redshift ($z>0.5$) absorbers are currently
poorly constrained. As suggested in section \ref{section_theory}, their
statistical properties can be probed by measuring the quantity $\Delta\langle
m_{obs}\rangle_\lambda-\Delta\langle m_{obs}\rangle_{\lambda'}$ for different
wavelengths (M\'enard et al., in preparation).

\section{Conclusion}

We have proposed a method aimed at detecting and interpreting gravitational
lensing by absorber systems. Whereas earlier studies focused on strong
Lyman-$\alpha$ systems, we suggest to use metal absorbers because they allow
for the compilation of larger samples and they can be detected from the ground
over favorable redshift ranges for lensing.
Due to magnification effects, a given population of absorbers will give rise
to a distribution of \emph{induced} quasar magnitude changes, which is
conveniently described by its centered moments $\langle \delta m^i\rangle$.
The corresponding magnitude changes that can be observed,
$\Delta\langle m_{obs}^i\rangle=\left\langle{m^i_\mathrm{lens}}\right\rangle-
\left\langle{m^i_\mathrm{ref}}\right\rangle$, depend on the shape of the quasar
luminosity function and the completeness of the survey.  We have shown how the
observed magnitude changes $\Delta\langle m_{obs}^i\rangle$ can be used to
recover the induced ones $\langle \delta m^i\rangle$ and then constrain the
mass distribution of the absorbing galaxies.
As a result of the magnification bias, lensing effects can effectively
increase or decrease the brightness of the detected sources, depending on
their luminosity function.  Furthermore if the sources exhibit a power law
distribution of luminosities, lensing effects due to absorbers \emph{cannot}
be observed, whatever their amplitude.

Our method can be applied to various absorption lines (MgII, FeII, SiII, CIV,
etc.). In order to explore its feasibility, we have computed the expected
lensing effects for a population of MgII absorbers.  Using the observed
distribution of rest equivalent widths as a function of impact parameter
measured by Steidel \e (1994), and assuming that the mass profile of absorber
systems is described by that of an isothermal sphere with a mean velocity
dispersion $100<\langle\sigma_v\rangle<200$ km
s$^{-1}$, we have shown that:\\
\indent $\bullet~$ Quasars behind MgII absorbers with
$0.3<W_0<1.5~\mathrm{\AA}$ are in average brighter by $\left\langle \delta
  m\right\rangle=-0.03$ to $-0.1$ magnitude. For systems with
$1.5<W_0<3.2~\mathrm{\AA}$ these values reach $-0.1$ to $-0.4$ magnitude.\\
\indent $\bullet~$ The corresponding scatter of magnitude increases by
$\left\langle \delta m^2\right\rangle=0.01$ to $0.15$ if the gas distribution
follows that of the dark matter, and $0.01$ to $0.25$ if we
allow for some scatter between these two quantities.\\
\indent Using the quasar luminosity distribution based on the 2dF Quasar
survey (Croom \e\ 2004) and realistic completeness functions we have shown
that:\\ 
\indent $\bullet~$ the corresponding magnitude changes that can be observed
are $\Delta \langle m_{obs}\rangle=-0.01$ to $-0.04$ for systems with
$0.3<W_0<1.5~\mathrm{\AA}$, and $-0.03$ to $-0.14$ in the
range $1.5<W_0<3.2~\mathrm{\AA}$.\\
\indent $\bullet~$ The corresponding variances in observed magnitudes
change by $\Delta \langle m_{obs}^2\rangle=-0.0005$ to $0.06$.
These values turn out to be very sensitive to the shape of the relevant
completeness
function.

Whereas our model provides us with estimates of the unavoidable magnitude
changes $\langle \delta m^i\rangle$ due to magnification, it does not allow us
to determine, in the general case, the detectability of $\Delta\langle
m_{obs}^i\rangle$ since effects of extinction by dust need to be included.
Observing at sufficiently large wavelengths can however isolate the lensing
effects.  In this case, we have shown that if MgII absorbers trace galactic
halos, about 2000 quasars with such systems will provide us with a detection
of the mean magnitude shift due to gravitational lensing.  Such a measurement
will allow us to constrain the mean mass of absorbing galaxies.  Detecting a
change in the magnitude distribution variances requires about $10\,000$
quasars with absorbers.  However, using our method to probe the stochasticity
in the gas-dark matter relation will require next generation surveys.\\ In
this paper we have considered the absorbing galaxies to be isolated.  Due to
the intrinsic clustering of galaxies, neighoring structures, that do not
necessarily give rise to absorption lines, might increase the magnification
effects. Quantifying this effect will require further investigations.

In addition to probing the mass distribution of absorber systems, our analysis
suggests that quasars behind metal absorbers are not necessarily dimmed, as is
usually mentioned in the literature, but they can actually be brighter if
extinction effects are smaller than the gravitational magnification effects
presented above. The claim that magnitude limited surveys might miss a number
of quasars with strong absorbers must therefore be revisited.  Applying our
method to large surveys will allow us to quantify this effect.

\section*{Acknowledgements}

I thank John Bahcall, Neal Dalal, Gil Holder, Alison Farmer and Matthias
Bartelmann for useful discussions. I acknowledge the Florence Gould foundation
for its financial support.

   \end{document}